\documentclass[aps,prl,twocolumn,groupedaddress,floatfix,showpacs]{revtex4}
\usepackage{graphicx}
\begin{document}
\title{Near-Field Modification of Interference Components in Surface Plasmon Resonance}
\author{Robert P. Schumann and Stephen Gregory}
\affiliation{Department of Physics, Oregon Center for Optics\\
and\\
Materials Science Institute\\ 
University of Oregon\\ 
Eugene, OR 97403-1274} 
\begin{abstract}
In an attenuated total reflection (ATR) geometry, energy extracted from an incident light
beam by surface plasmon polariton (SPP) creation is implicitly considered to dissipate as
heat. However, we show that a sharpened tip interacting with the SPP evanescent field
can redirect some of the energy before such dissipation occurs. This behavior is
examined with simple computer simulations and an analogy is drawn between it and a
modified Mach-Zehnder interferometer. We also discuss "interaction-free"
measurements with our set-up.
\end{abstract}
\maketitle
For a metal film in an ATR geometry, a minimum in the reflected light intensity at a specific
angle is usually viewed as a signature of surface plasmon resonance (SPR). While this no
doubt is the case, all that can \emph{directly} be concluded is that there is some energy
``missing'' from the specular beam. As momentum conservation prevents light
transmission beyond the film (if it is smooth) one infers that the missing energy
corresponding to the minimum is deposited in the film as heat, and this can indeed be
readily detected. \cite{inagaki81}

On the other hand, the standard analysis of the ATR geometry \cite{raether86} is not
specifically concerned with energy, but instead considers the interference of light
reflected
from the front and back surfaces of the film and reproduces the experimentally-observed
minimum.  As with all interference phenomena, the existence of a minimum in one
interference outcome implies a maximum in a complementary outcome. Reconciling the
energy and interference pictures indicates that the missing energy must be carried into
the complementary outcome as SPPs and eventually converts to heat. However, this
sequence is not necessarily inevitable, because one may imagine intercepting some of
this energy and redirecting it - essentially changing the destiny of the SPPs. To realize
this situation we use the tip of an apertureless near-field scanning optical microscope
(A-NSOM) to redirect
some SPP energy and examine the effect of this on the specular minimum. When
specifically studying SPPs this A-NSOM variant is referred to as a scanning plasmon
near-field microscope (SPNM). \cite{specht92,kim_yk95}) 
\begin{figure}
\includegraphics[scale=0.4]{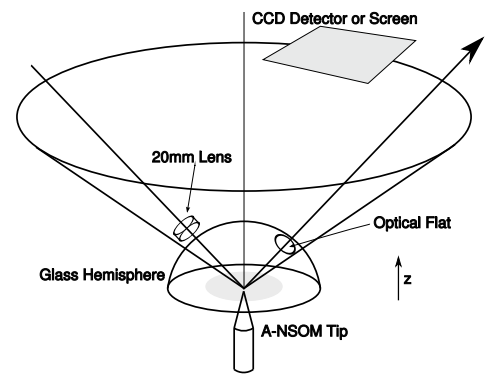}
\caption{Experimental arrangement for observing specular reflection and SPP ring
\label{fig:1}}
\end{figure}

In the Kretschmann-Raether (K-R) \cite{kretschmann71,raether86} ATR geometry light is
incident through a dielectric prism, on the hypotenuse of which a thin metal film is
deposited. (See Fig. \ref{fig:1}.) Our SPNM employs a hemispherical glass prism and
maintains either a vacuum or inert gas atmosphere ("ambient") over the metal film
surface, allowing us to take
advantage of the optimum performance of silver as a material for studying SPPs. The
scanning probe is tungsten, which does not itself support SPP modes and can be etched
to a very sharp tip. Incident light with a wavelength of $632.8\mbox{nm}$ is focused on
the silver film
with a $20\mbox{mm}$ lens, in combination with the $25.4\mbox{mm}$ radius surface
of the prism. The calculated beam waist is approximately $3\mu\mbox{m}$, giving an
approximately $5\mu\mbox{m}\times3\mu\mbox{m}$ elliptical spot at the metal film.

It is important that our incident beam is focused, but initially let us consider the
reflection of a plane wave in a dielectric-metal-dielectric structure. Solution of the
Fresnel Equations for this multilayer system gives the following expression for the
reflectivity:
\begin{equation}
r_{123}=\frac{r_{12}+r_{23}\exp\left(2ik_{2z}d\right)}{1+r_{12}r_{23}
\exp\left(2ik_{2z}d\right)}
\label{eqn:1}
\end{equation}
The first dielectric (the prism in our case) is medium 1, the metal film is medium 2 and
the ambient is medium 3. $d$ is the film thickness and $k_{2z}$ is the component of the
wave vector in medium 2 perpendicular to the interfaces. Writing the reflectivity in this
form emphasizes that the numerator is the sum of the reflections $r_{12}$ from the
glass/metal interface and $r_{23}$ from the metal/ambient interface, the latter being
modified by the factor $\exp\left(2ik_{2z}d\right)$. As $k_{2z}$ has a large imaginary
part, this factor decreases exponentially with film thickness. Above the critical angle
$r_{12}$ varies only slightly with angle of incidence; $r_{23}$, on the other hand, is
peaked and because it shifts into antiphase with $r_{12}$ at the ``resonance angle'',
$r_{123}$ displays a minimum at that angle, the depth of which is non-monotonic in $d$.

\begin{figure}
\includegraphics[scale=0.4]{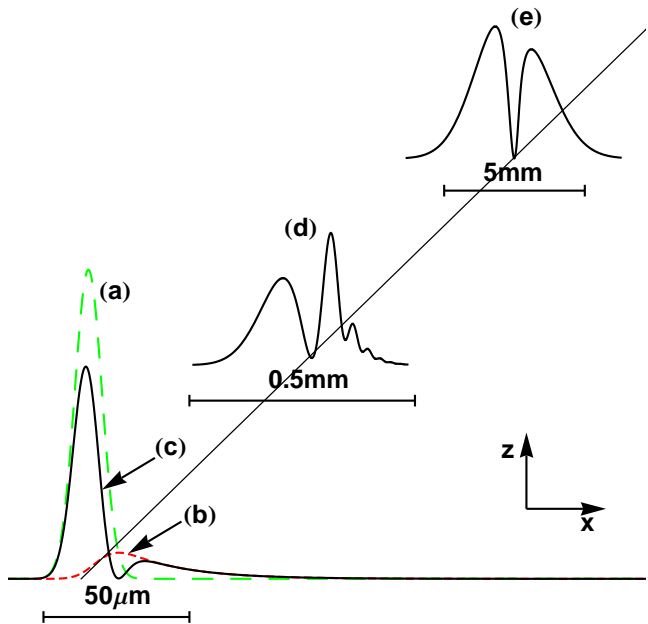}
\caption{(Color online) Propagation of the specular beam. Distance is measured from the
incident Gaussian center at the 1-2 interface ($0\mbox{mm}$) along the direction of the
SPR far-field minimum. Intensities of: (a) the component reflected from the 1-2 interface,
(b) the component reflected from the 2-3 interface and (c) the summed 1-2 and
2-3 interface reflected components, all evaluated at $0\mbox{mm}$; (d) the propagating
sum at $6.1\mbox{mm}$; (e) the propagating sum at $10.1\mbox{mm}$. Scale bars
indicate
increasing width of the beam profile.
\label{fig:2}}
\end{figure}

In the case of an incident focused Gaussian beam the above approach can be extended 
by wave-vector decomposition of the incident
field  \cite{chen_wp76,shah83,chuang86} and
calculation of the reflected fields. Fig. \ref{fig:2} shows spatial profiles of the separate
intensities reflected from the 1-2 and 2-3 interfaces as well as the summed beam
intensity.  As the sum propagates in medium 1 it exhibits oscillations at distances
intermediate between the near- and the far-field. Interestingly, however, these are
present in the intensity of $r_{23}$ alone. Similar oscillations have been predicted and
observed for an ATR apparatus involving two-surface "long-range" surface plasmons.
\cite{andaloro05,simon07}.  It is possible, with an appropriate lens, to retain oscillations
in the far-field. Since the spherical surface of the prism also has this effect, to obtain a
far-field beam free of oscillations we polish a small optical flat into the prism
(see Fig. \ref{fig:1}) and the expected ``notched Gaussian'' is then observed.

Although there is no propagating solution for the field in medium 3, there exists a formal
solution which decays exponentially in the negative z-direction; i.e. it is evanescent. The
x-dependent profile of this solution is identical to that of $r_{23}$, so its intensity is not
shown in Fig. \ref{fig:2}.  It can, however, be identified with the SPP field propagating
and decaying along the interface. 

The above treatment is necessary to produce details of the focused Gaussian beam
situation, but for much of the following discussion we can refer to the simpler plane-wave
expression, Eq. (\ref{eqn:1}). From this we can see that there is a critical thickness
$d_{0}$, defined by $r_{23}\exp\left(2ik_{2z}d\right)=-r_{12}$ at which the notch in the
specular beam  reaches maximum area (and zero intensity). Suppose
now that, all else remaining the same, $r_{23}$ is somehow reduced to $\alpha r_{23}$, 
where $\alpha<1$. If the film thickness
$d<d_{0}$, the notch area in the specular beam is \emph{increased}, which might seem
surprising, if viewed in terms of a ``stronger SPR''. However, it is quite consistent with an
energetic perspective: The notch is larger because less energy is present in
the specular beam, and this accords with any mechanism that reduces
$r_{23}$ by dissipation or scattering. On the other hand, if  $\alpha<1$ and $d>d_{0}$,
the notch area is \emph{decreased}. Although more energy is present in the specular
beam, more energy also is dissipated or scattered by the mechanism
reducing $r_{23}$. The source of the latter energy can only be SPP energy that otherwise
would eventually have been "lost" as heat.

We investigate the cases contrasted above by modifying the $r_{23}$ field with a SPNM 
tip. In these runs, the tip is moved only in the z-direction. However, we also
perform STM and A-NSOM-type scans to judge the quality of our deposited films.  If the
incident beam is focused to a small spot, the tip can interact with a substantial portion of
the SPP field. (It will also interact with the non-resonant components of the
evanescent field of the incident beam, but to a much lesser extent, dependent on the
film thickness.) The main effect of bringing the tip near the metal surface is to elastically
scatter SPPs within the plane of the interface. \cite{specht92} These can subsequently
decay into an expanding hollow cone of radiation \cite{kim_yk95} that can be detected
outside the hemisphere with a CCD camera. The light either impinges directly on the
CCD detector or, if the image is too large, onto a translucent screen, which can then
be imaged.
\begin{figure}
\includegraphics[scale=0.65]{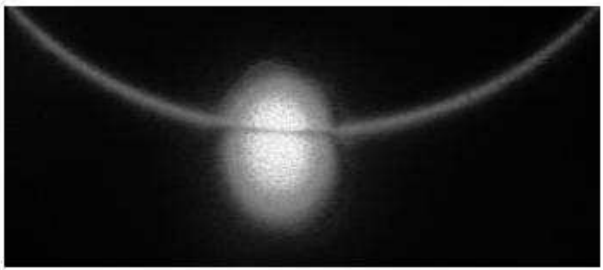}
\caption{Specular reflection spot and portion of SPP ring. To show the ring most clearly,
the tip is within tunneling distance of the surface. \label{fig:3}}
\end{figure}

\begin{figure}
\includegraphics[scale=0.40]{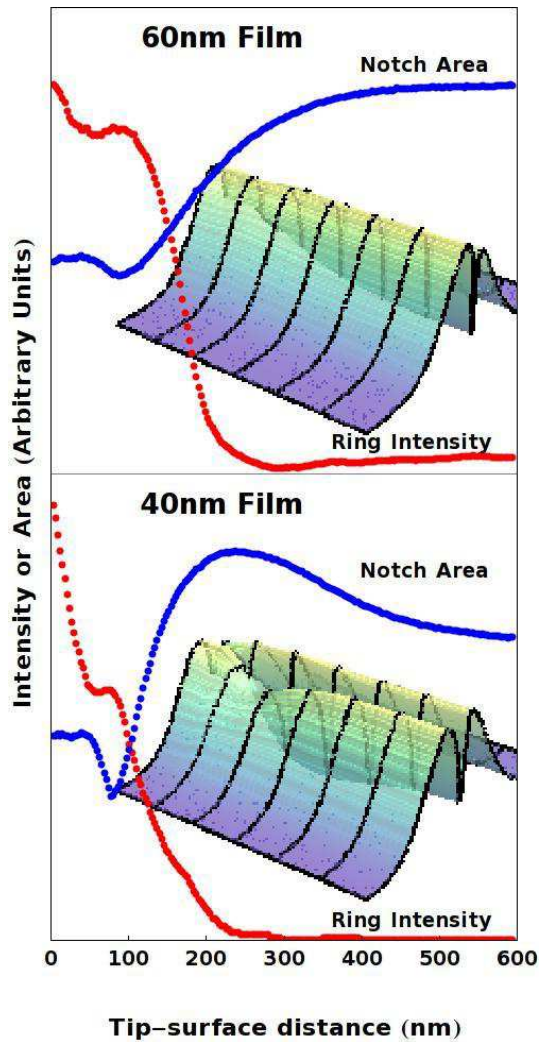}
\caption{(Color online) Dependence of the area of the specular notch and the total ring
intensity on the tip-surface distance for 40nm and 60nm silver films. Inset surfaces
are the profiles of the specular spot as a function of tip-surface distance.
\label{fig:4}}
\end{figure}

Fig \ref{fig:3} is an image of part of the circular conic section (``SPP ring'') and the
specular Gaussian beam. Note that the notch is only approximately centered, the exact
position within the Gaussian profile depends on the angle of incidence. The ring
intensity and notch profile calculated from the appropriate pixel values are plotted
in Fig \ref{fig:4} against separation of the tip from the metal surface.
Data for a $40\mbox{nm}$ and $60\mbox{nm}$ film are shown. At a wavelength
of $632.8\mbox{nm}$, the critical thickness of a silver film is calculated to
be $d_0 \approx 55\mbox{nm}$.  The notch area is calculated as the area between the
specular beam profile and a Gaussian fit based on the outer regions of the profile.

For the $40\mbox{nm}$ film, at large tip-surface separation
the notch area increases as the
tip approaches the surface, but initially the ring intensity does not increase. It is possible
that, for large separation, the tip just dissipates energy (as heat in the tip) and does not
appreciably scatter SPPs. Modeling this behavior by adding a small imaginary part to the
dielectric constant of medium 3  which depends inversely on separation reproduces the
observed behavior for $r_{123}$.  At a separation of about $250\mbox{nm}$ the ring
intensity begins to increase, indicating that the tip is elastically scattering SPPs. The
notch area now begins to decrease, again in agreement with the expected behavior of
$r_{123}$ as $\left|\alpha r_{23}\exp\left(2ik_{2z}d\right)\right|$ becomes less than
$\left|r_{12}\right|$. The ring intensity continues to rise and the notch area continues
to decrease until the separation reaches about $70\mbox{nm}$. Here the notch exhibits
a sudden phase and zenith angle shift, the area rising slightly. There is a matching
feature in the ring. With smaller separation the ring intensity continues to rise, but the
notch area remains fairly constant. 

For the $60\mbox{nm}$ film,  the notch area decreases monotonically
with decreasing tip-surface separation, but as with the $40\mbox{nm}$ film and for the
same reason, the ring intensity does not increase until about $250\mbox{nm}$. (Above
this separation the ring intensity actually seems to be non-zero and decreasing slightly,
but this is probably an artifact.) With continued decrease of the tip-surface separation,
the notch area decreases monotonically. At about $250\mbox{nm}$ the ring intensity
begins increasing rapidly, as with the $40\mbox{nm}$ film, indicating that elastic SPP
scattering is now occuring. At present, we do not understand the significance of the
$250\mbox{nm}$ separation. We note, however, that this same value was obtained
in Ref. \onlinecite{kim_yk95}. When the separation reaches about $90\mbox{nm}$
there are features in both the notch area and ring intensity similar to those for
the $40\mbox{nm}$ film. 

The area of the notch for small separation is rather important: In principle, the total
energy available for production of either the SPP ring or heat is that which is missing
from the specular beam i.e. the notch. If the tip could somehow scatter \emph{all} that
notch energy
into ring energy then the energy corresponding to the minimum notch area would
be equal to that ring energy. In practice, however, scattered SPPs can later decay into
heat and the scattering efficiency of the tip depends on the tip shape and distance of
the tip ``downstream'' from the center of the incident beam. Thus, the small-separation
notch area is larger than the limiting minimum value. It is interesting that, even after the
notch area becomes constant, the overall efficiency of ring light production continues
to rise at small separations - quite dramatically in the case of the $40\mbox{nm}$  film.

\begin{figure}
\includegraphics[scale=0.6]{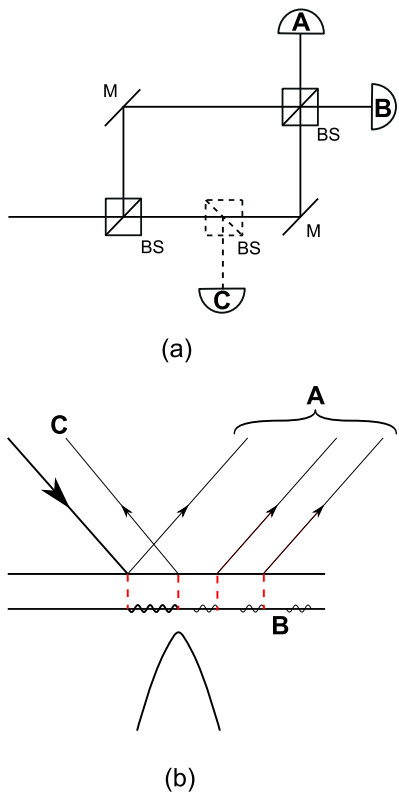}
\caption{(Color online) (a) Mach-Zehnder arrangement with extra beam-splitter. A, B and
C are light detectors, BS and M denote beam splitters and mirrors respectively.
(b) Schematic of the ATR arrangement with movable tip.
\label{fig:5}}
\end{figure}

We point out that our SPNM experiment is analogous to the modified
Mach-Zehnder interferometer shown in Fig. \ref{fig:5}a. Absent the extra (dashed)
beamsplitter, the lengths of the arms and split fractions are adjusted such that no light
reaches detector A. In the SPNM, the partial reflection off the 1-2 (glass-metal) interface
and the partial conversion to a propagating SPP wave along the 2-3
(metal-ambient) interface are analogous to the effect of the first beamsplitter in the
Mach-Zehnder interferometer. A difference is that the SPNM produces a light
wave and a SPP wave which, in a particle picture, would be viewed as photon-SPP
superposition. Subsequently, the reflected wave propagates out to the far field, which
we analogize to traversal of the upper arm in the interferometer. The SPP wave, on the
other hand travels, in the absence of the tip, an exponentially-distributed distance along
the 1-2 interface and tunnels to a wave in the glass which also then propagates out to the
far field. The observation of the notch produced by the sum of the two outgoing waves
at "A" in the far field is analogous to the observation of no light at detector A in the
interferometer, (although the notch is not a perfect null in our experiments). We
identify the distributed conversion of the SPPs to heat with the output to detector B in
the interferometer.

As already discussed, the presence of the tip (except for the large separation range of
the $40\mbox{nm}$ film) causes an increase in the light detected in the notch while also
causing light to be detected in the SPP ring. In the two-dimensional representation of
Fig. \ref{fig:5}b this is indicated by C, which can actually be anywhere in the ring. In the
same way, the introduction of another beamsplitter into the lower arm of the
interferometer causes light to be measured at detector A while also producing light at
detector C.  We have previously argued that, in our experiment,  the presence of the tip
redirects some energy that would have dissipated into heat, and which we have labelled
"B" in Fig. \ref{fig:5}b. Similarly, in the interferometer the introduction of the extra
beamsplitter has the effect of decreasing the light measured by detector B. Note that
advancing the tip in the SPNM experiment has an interesting advantage over the
introduction of a beamsplitter in the interferometer: There is no extra phase shift in the
former situation because the tip interacts with an evanescent SPP field, and so the new
element does not unbalance the interferometer.

When the input of the SPMN is a greatly attenuated (single-photon) source and we
photon count, collecting light with a multimode fiber from the notch minimum, the
arrangement is essentially equivalent to that in the Elitzur-Vaidman (EV) "bomb-detector
problem" \cite{elitzur93} and its experimental realization. \cite{kwiat95a} In our
experiment, the increased rate of
photon detection (in principle from zero, which can be approached for a film of thickness
$d_0$),  is an "interaction-free measurement" of the presence of the tip. On the other
hand, photons detected at C (i.e. in the ring) are essentially equivalent to
an explosive detection of the EV bomb and are always anticoincident with those at A. We
continue to explore the consequences of this kind of detection for near-field microscopy. 

Sponsored by ONR through the Oregon Nanoscience and Microtechnologies Institute.

\begin{thebibliography}{12}
\expandafter\ifx\csname natexlab\endcsname\relax\def\natexlab#1{#1}\fi
\expandafter\ifx\csname bibnamefont\endcsname\relax
  \def\bibnamefont#1{#1}\fi
\expandafter\ifx\csname bibfnamefont\endcsname\relax
  \def\bibfnamefont#1{#1}\fi
\expandafter\ifx\csname citenamefont\endcsname\relax
  \def\citenamefont#1{#1}\fi
\expandafter\ifx\csname url\endcsname\relax
  \def\url#1{\texttt{#1}}\fi
\expandafter\ifx\csname urlprefix\endcsname\relax\def\urlprefix{URL }\fi
\providecommand{\bibinfo}[2]{#2}
\providecommand{\eprint}[2][]{\url{#2}}

\bibitem[{\citenamefont{Inagaki et~al.}(1981)\citenamefont{Inagaki, Kagami, and
  Arakawa}}]{inagaki81}
\bibinfo{author}{\bibfnamefont{T.}~\bibnamefont{Inagaki}},
  \bibinfo{author}{\bibfnamefont{K.}~\bibnamefont{Kagami}}, \bibnamefont{and}
  \bibinfo{author}{\bibfnamefont{E.~T.} \bibnamefont{Arakawa}},
  \bibinfo{journal}{Phys. Rev. B} \textbf{\bibinfo{volume}{24}},
  \bibinfo{pages}{3644} (\bibinfo{year}{1981}).

\bibitem[{\citenamefont{Raether}(1986)}]{raether86}
\bibinfo{author}{\bibfnamefont{H.}~\bibnamefont{Raether}},
  \emph{\bibinfo{title}{Surface Plasmons on Smooth and Rough Surfaces and on
  Gratings}} (\bibinfo{publisher}{Springer-Verlag, Berlin},
  \bibinfo{year}{1986}).

\bibitem[{\citenamefont{Specht et~al.}(1992)\citenamefont{Specht, Pedarnig,
  Heckel, and H\"{a}nsch}}]{specht92}
\bibinfo{author}{\bibfnamefont{M.}~\bibnamefont{Specht}},
  \bibinfo{author}{\bibfnamefont{J.~D.} \bibnamefont{Pedarnig}},
  \bibinfo{author}{\bibfnamefont{W.~M.} \bibnamefont{Heckel}},
  \bibnamefont{and} \bibinfo{author}{\bibfnamefont{T.~W.}
  \bibnamefont{H\"{a}nsch}}, \bibinfo{journal}{Phys. Rev. Lett.}
  \textbf{\bibinfo{volume}{68}}, \bibinfo{pages}{476} (\bibinfo{year}{1992}).

\bibitem[{\citenamefont{Kim et~al.}(1995)\citenamefont{Kim, Lundquist,
  Helfrich, Mikrut, Wong, Auvil, and Ketterson}}]{kim_yk95}
\bibinfo{author}{\bibfnamefont{Y.-K.} \bibnamefont{Kim}},
  \bibinfo{author}{\bibfnamefont{P.~M.} \bibnamefont{Lundquist}},
  \bibinfo{author}{\bibfnamefont{J.~A.} \bibnamefont{Helfrich}},
  \bibinfo{author}{\bibfnamefont{J.~M.} \bibnamefont{Mikrut}},
  \bibinfo{author}{\bibfnamefont{G.~K.} \bibnamefont{Wong}},
  \bibinfo{author}{\bibfnamefont{P.~R.} \bibnamefont{Auvil}}, \bibnamefont{and}
  \bibinfo{author}{\bibfnamefont{J.~B.} \bibnamefont{Ketterson}},
  \bibinfo{journal}{Appl. Phys. Lett.} \textbf{\bibinfo{volume}{66}},
  \bibinfo{pages}{3407} (\bibinfo{year}{1995}).

\bibitem[{\citenamefont{Kretschmann}(1971)}]{kretschmann71}
\bibinfo{author}{\bibfnamefont{E.}~\bibnamefont{Kretschmann}},
  \bibinfo{journal}{Z. Phys.} \textbf{\bibinfo{volume}{241}},
  \bibinfo{pages}{313} (\bibinfo{year}{1971}).

\bibitem[{\citenamefont{Chen et~al.}(1976)\citenamefont{Chen, Ritchie, and
  Burstein}}]{chen_wp76}
\bibinfo{author}{\bibfnamefont{W.~P.} \bibnamefont{Chen}},
  \bibinfo{author}{\bibfnamefont{G.}~\bibnamefont{Ritchie}}, \bibnamefont{and}
  \bibinfo{author}{\bibfnamefont{E.}~\bibnamefont{Burstein}},
  \bibinfo{journal}{Phys. Rev. Lett.} \textbf{\bibinfo{volume}{37}},
  \bibinfo{pages}{993} (\bibinfo{year}{1976}).

\bibitem[{\citenamefont{Shah and Tamir}(1983)}]{shah83}
\bibinfo{author}{\bibfnamefont{V.}~\bibnamefont{Shah}} \bibnamefont{and}
  \bibinfo{author}{\bibfnamefont{T.}~\bibnamefont{Tamir}}, \bibinfo{journal}{J.
  Opt. Soc. Am.} \textbf{\bibinfo{volume}{73}}, \bibinfo{pages}{37}
  (\bibinfo{year}{1983}).

\bibitem[{\citenamefont{Chuang}(1986)}]{chuang86}
\bibinfo{author}{\bibfnamefont{S.~L.} \bibnamefont{Chuang}},
  \bibinfo{journal}{J. Opt. Soc. Am. A} \textbf{\bibinfo{volume}{3}},
  \bibinfo{pages}{593} (\bibinfo{year}{1986}).

\bibitem[{\citenamefont{Andaloro et~al.}(2005)\citenamefont{Andaloro, Deck, and
  Simon}}]{andaloro05}
\bibinfo{author}{\bibfnamefont{R.~V.} \bibnamefont{Andaloro}},
  \bibinfo{author}{\bibfnamefont{R.~T.} \bibnamefont{Deck}}, \bibnamefont{and}
  \bibinfo{author}{\bibfnamefont{H.~J.} \bibnamefont{Simon}},
  \bibinfo{journal}{J. Opt. Soc. Am. B} \textbf{\bibinfo{volume}{22}},
  \bibinfo{pages}{1512} (\bibinfo{year}{2005}).

\bibitem[{\citenamefont{Simon et~al.}(2007)\citenamefont{Simon, Andaloro, and
  Deck}}]{simon07}
\bibinfo{author}{\bibfnamefont{H.~J.} \bibnamefont{Simon}},
  \bibinfo{author}{\bibfnamefont{R.~V.} \bibnamefont{Andaloro}},
  \bibnamefont{and} \bibinfo{author}{\bibfnamefont{R.~T.} \bibnamefont{Deck}},
  \bibinfo{journal}{Opt. Lett.} \textbf{\bibinfo{volume}{32}},
  \bibinfo{pages}{1590} (\bibinfo{year}{2007}).

\bibitem[{\citenamefont{Elitzur and Vaidman}(1993)}]{elitzur93}
\bibinfo{author}{\bibfnamefont{A.~C.} \bibnamefont{Elitzur}} \bibnamefont{and}
  \bibinfo{author}{\bibfnamefont{L.}~\bibnamefont{Vaidman}},
  \bibinfo{journal}{Found. Phys.} \textbf{\bibinfo{volume}{23}},
  \bibinfo{pages}{987} (\bibinfo{year}{1993}).

\bibitem[{\citenamefont{Kwiat et~al.}(1995)\citenamefont{Kwiat, Weinfurter,
  Herzog, Zeilinger, and Kasevich}}]{kwiat95a}
\bibinfo{author}{\bibfnamefont{P.~G.} \bibnamefont{Kwiat}},
  \bibinfo{author}{\bibfnamefont{H.}~\bibnamefont{Weinfurter}},
  \bibinfo{author}{\bibfnamefont{T.}~\bibnamefont{Herzog}},
  \bibinfo{author}{\bibfnamefont{A.}~\bibnamefont{Zeilinger}},
  \bibnamefont{and} \bibinfo{author}{\bibfnamefont{M.}~\bibnamefont{Kasevich}},
  \bibinfo{journal}{Phys. Rev. Lett.} \textbf{\bibinfo{volume}{74}},
  \bibinfo{pages}{4763} (\bibinfo{year}{1995}).

\end{thebibliography}
\end{document}